\documentclass[conference]{IEEEtran}

\usepackage{cite}
\usepackage{amsmath,amssymb,amsfonts}
\usepackage{algorithmic}
\usepackage{graphicx}
\usepackage{textcomp}

\usepackage{graphicx}
\usepackage{amsmath}
\usepackage{color}
\usepackage{soul}
\usepackage{cite}
\usepackage[caption=false,font=footnotesize]{subfig}
\usepackage{enumerate} 
\usepackage{hyphenat}
\usepackage[numbers,sort&compress]{natbib}
\usepackage{hyperref}
\usepackage{amssymb}
\usepackage{textcomp}

\usepackage{url}
\usepackage{array} 
\usepackage{multirow}
\usepackage{longtable}
\usepackage[utf8]{inputenc}

\def\BibTeX{{\rm B\kern-.05em{\sc i\kern-.025em b}\kern-.08em
    T\kern-.1667em\lower.7ex\hbox{E}\kern-.125emX}}

\ifCLASSINFOpdf

\else
 
\fi

\begin{document}

\title{Predicting Succinylation Sites in Proteins with Improved Deep Learning Architecture}

\author{\IEEEauthorblockN{Olusola T. Odeyomi and Gergely Zaruba}
\IEEEauthorblockA{Department of Electrical Engineering and Computer Science\\
Wichita State University\\
Wichita, Kansas 67260--0083\\
Email: otodeyomi@shockers.wichita.edu\\
gergely.zaruba @wichita.edu}
\and
}

\maketitle

\begin{abstract}
Post-translational modifications (PTMs) in proteins occur after the process of translation. PTMs account for many cellular processes such as deoxyribonucleic acid (DNA) repair, cell signaling and cell death. One of the recent PTMs is succinylation. Succinylation modifies lysine residue from $-1$ to $+1$. Locating succinylation sites using experimental methods, such as mass spectrometry is very laborious. Hence, computational methods are favored using machine learning techniques. This paper proposes a deep learning architecture to predict succinylation sites. The performance of the proposed architecture is compared to the state-of-the-art deep learning architecture  and other traditional machine learning techniques for succinylation. It is shown from the performance metrics that the proposed architecture provides a good trade-off between speed of computation and classification accuracy.
\end{abstract}


\IEEEpeerreviewmaketitle

\section{Introduction}
Post-translational modification (PTMs) occur in proteins by the addition of a covalent bond after protein is synthesized from messenger RNA (mRNA) by ribosomes. PTMs are very important in many cellular processes such as DNA repair, cell death and cell signaling. Sites in protein that undergo PTMs are those that can act as a nucleophile in reactions. There are many types of PTMs such as phosphorylation, glycosylation, succinylation etc. Protein succinylation is a recently emerged protein post-translational modification on lysine. Succinylation introduces a large structural moeity (e.g., 100 Da) and its modifies the net charge of modified lysine residue from -1 to +1. It is suggested that protein succinylation is important in cellular metabolism and many other cellular functions. Experimental methods for determining protein succinylation sites, such as mass spectrometry, are laborious and not cost-effective. Computational methods are thus favored.

Most existing work that uses computational methods employs machine learning techniques such as in iSuc-PseAAC \cite{xu2015isuc}, iSuc-PseOPT \cite{jia2016isuc}, pSuc-Lys \cite{jia2016psuc}, SuccinSite \cite{hasan2016succinsite}, SuccinSite 2.0 \cite{hasan2017systematic}, GPSuc \cite{hasan2018gpsuc}, and PSuccE \cite{ning2018detecting}. The  problem with traditional machine learning methods is the requirement for manual extraction of features. Also, traditional machine learning methods cannot handle very large FASTA file reliably. Most deep learning methods in literature are applied to phosphorylation rather than succinylation, and use one-hot encoding that does not account for the proper classification of features. However, there is a recent work that uses a deep learning architecture for succinylation with state-of-the-art performance. The authors called this architecture deepSuccinylSite-embedding \cite{thapa2020deepsuccinylsite}.

Thus, in this paper, the performance of deepSuccinylSite-embedding is compared with the performance of a proposed deep learning architecture called DeepSuccinylCNN+LSTM. While deepSuccinylSite uses 2D convolutional network, the novel architecture uses 1D Convolutional network + LSTM (Long Short-Term Memory). The motivation for this architecture is comparable performance  with deepSuccinylSite but with faster computational speed. This is very important when dealing with large FASTA file.

\section{Dataset Preprocessing}
The dataset for this computation was obtained from experimentally derived lysine residue reported in \cite{hasan2018gpsuc} and \cite{ning2018detecting}. Proteins with more than thirty percent $30\%$ sequence identity were removed using CD-HIT. This left a remainder of $5009$ succinylation sites, and $53,542$ non-succinylation sites. Out of these,  $4755$ succinylation sites and $50,565$ non-succinylation sites were used as training set. This meant that $254$ succinylation sites and $2977$ non-succinylation sites were used as independent test set. However, $5$ out of the $4755$ positive sites were lost because they contained other characters. The datasets were balanced using undersampling. The final training set had $4750$ positive and $4750$ negative sites. The final independent test set contained $254$ positive and $254$ negative sites. The optimal window size was $33$. If left and right side were less than half the size of the window, then a pseudo-residue “-” was used to recover all positive sites. Table 1 shows the size of the positive and negative sites for the training and independent test sets.

\begin{table*}
\caption{Size of Training and Independent Test Set }
\centering
    \begin{tabular}{| c |c| c | }
    \hline
Dataset  &  Positive & Negative\\
\hline
 Training &  4750 & 4750\\
 \hline
 Independent Test & 254 & 254\\
 \hline

    \end{tabular}
     
    \label{table:2}
\end{table*}

\section{Input Modeling}
The novel deep learning architecture Conv1D+LSTM  used keras embedding layer similar to DeepSuccinylSite. The embedding layer came before the first convolutional layer. The embedding layer output was a 21-dimensional vector space. The embedding layer grouped co-occuring items together. Other types of sophisticated embedding techniques common in natural language processing could be used, such as Word2Vec or GloVe, but they often come with increased cost of computation. The training set was divided into $80\%$ training and $20\%$ validation set. Validation was done at every epoch to prevent overfitting. Checkpointer was used to select the optimal model from the epochs based on validation accuracy. The protein sequence dataset was in FASTA format. The input dimension to the convolutional neural network had a dimension of $33$ by $21$, since the window size was $33$, and the vector size of the output from the embedding size was $21$.
An advantage of using deep learning is the elimination of handcrafted features, which is often laborious. 

\section{Deep Learning Architecture}
The input to the first layer of the deep learning architecture is the embedding layer. DeepSuccinylSite used a Lambda layer after this embedding layer, but the Lambda layer was not included in the architecture used in this paper. Three convolutional layers were used with $64$, $128$ and $256$ number of neurons respectively. The hidden layers were for hierarchical feature extraction. Although, DeepSuccinylSite used  $2D$ convolutional layers instead of $1D$ convolutional layers for more feature extraction, it was observed that it slowed down the speed of the computation in comparison to the improvement obtained. Hence, $1D$ convolutional layer was used together with LSTM in DeepSuccinylCNN+LSTM. The kernel size used was $17$, since the PTM site lied in the $17^{th}$ position.
The use of this kernel size removed the need for paddling. Dropouts of $0.6$ were inserted between the convolutional layers to avoid overfitting. The activation function was linear rectifier (ReLu) because it could minimize overfitting and maximize the predictive power of the model. A single maxpool layer was used followed by the LSTM layer. Three dense layers with neurons size $256$, $768$ and $1024$ were used. Dropouts of $0.5$, $0.5$ and $0.25$ were used between the dense layers respectively. The optimization used was Adam optimization, which is an adaptive stochastic gradient optimization algorithm.
Adam optimization inherently possesses the advantages of adaptive gradient optimization and root-mean-square propagation. Also, binary cross entropy was used as the loss function because this is a binary classification problem. The final output layer contained 2 nodes and used softmax activation function.  Figure 1 is a snapshot of the parameters for the simulation.

\begin{figure}[t!]
\centering
\includegraphics[width=3in]{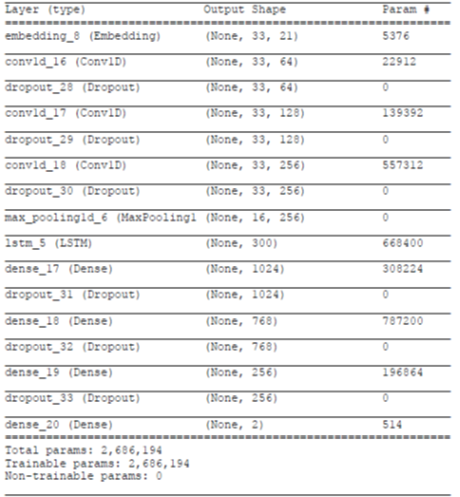}
\caption{A snapshot of the parameters for DeepSuccinylCNN+LSTM architecture}
\label{Drawing1}
\end{figure}

\section{Results}
For this simulation, $10-$fold cross validation was used. This means that the training set was divided into $9$ parts for training and $1$ part for validation at every epoch. This process is repeated, so that each part is used as the validation set. The performance metrics for the simulation are Matthew correlation, specificity and sensitivity. These performance metrics are defined below.
\subsection{Sensitivity}
Sensitivity is a measure of how many positive succinylation sites were correctly identified. It is given by the (3) below:
\begin{equation}
    sensitivity = \frac{TP}{TP + FN}
\end{equation}
where $TP$ refers to the True Positives and $FN$ refers to the False Negatives.
\subsection{Specificity}
Specificity measures the number of negative succinylation sites correctly identified. 
\begin{equation}
    specificity = \frac{TN}{TN + FP}
\end{equation}
where $TN$ refers to the True Negatives and $FP$ refers to the False Positives. The $TP$, $TN$, $FP$ and $FN$ were gotten from the confusion matrix.
\subsection{Matthew Correlation Coefficient}
The Matthew correlation coefficient (MCC) is a measure of the correlation between the predicted binary class and the observed binary class. It ranges from $-1$ to $+1$. A coefficient of $-1$ means that there is a total disagreement between the predicted and the observed binary classifications. A coefficient of $0$ means that the prediction is random. A coefficient of $1$ means that there is a perfect agreement between the predicted and the observed binary classifications. The formula for MCC is given as 
\begin{equation}
    MCC = \frac{(TP\times TN) - (FP \times FN)}{\sqrt{(TP+FP)(TP+FN)(TN+FP)(TN+FN)}}
\end{equation}
\subsection{Receiver Operating Characteristics Curve}
The receiver operating characteristic curve (ROC) is a graphical tool to investigate the discriminatory ability of a binary classifier. It compares sensitivity to specificity, with the true positive rate on the y-axis, and the false positive rate on the x-axis. The diagonal line in the plot divides the ROC space. Points above this diagonal line depicts good classification results, while points below this line depicts poor classification results.
\subsection{Area Under ROC Curve}
The area under the ROC curve (AUC) measures the entire area area under the ROC curve. It can be defined as the probability that a binary classifier will rank higher a randomly chosen positive site than a randomly chosen negative site. A classifier whose predictions are completely wrong has an AUC of $0.0$, while a classifier whose predictions are completely correct has an AUC of $1.0$. Thus, the AUC ranges from $0$ to $1$.

The simulation used the same datasets, window size and embedding size to compare the performance of the state-of-the-art DeepSuccinylSite deep learning architecture and the proposed deep learning architecture.  Both deep learning architectures were trained on the training set, but their performance were measured on the independent test set. The performance of both deep learning architectures is shown in Table II.

\begin{table*}
\caption{DeepSuccinylSite Versus DeepSuccinylCNN+LSTM}
\centering
    \begin{tabular}{c |c| c |c}
    \hline
Prediction Schemes  &  Sensitivity & Specificity & MCC\\
\hline
 DeepSuccinylSite &  0.73 & 0.70 & 0.43\\
 DeepSuccinylSiteCNN+LSTM & 0.77 & 0.63 &0.39\\

\hline
    \end{tabular}
     
    \label{table:2}
\end{table*}
The ROC curves for DeepSuccinylSite and DeepSuccinylCNN+LSTM are both shown in Figs. 2 and 3 respectively. 
\begin{figure}[t!]
\centering
\includegraphics[width=3in]{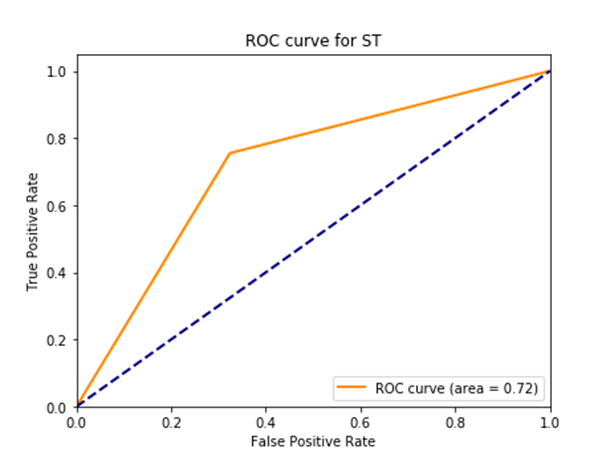}
\caption{ROC curve for DeepSuccinylSite.}
\label{Drawing1}
\end{figure}
\begin{figure}[t!]
\centering
\includegraphics[width=3in]{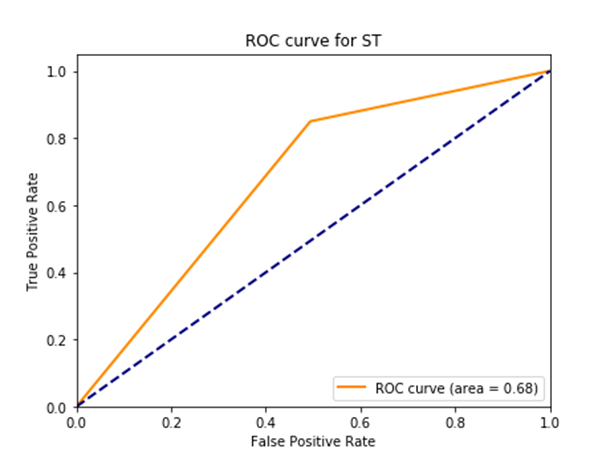}
\caption{ROC curve for DeepSuccinylCNN+LSTM.}
\label{Drawing1}
\end{figure}
It can be observed from Table II that DeepSuccinylSite has a higher MCC and specificity than DeepSuccinylCNN+LSTM, but a lower sensitivity. However, the difference is not much. Also, from Figs. 2 and 3, it can also be observed that DeepSuccinylSite has a higher AUC than DeepSuccinylCNN+LSTM. The main advantage of DeepSuccinylCNN+LSTM over DeepSuccinylSite is faster computation. The average time for running an epoch in DeepSuccinylSite was 29 seconds. The average time for running an epoch in DeepSuccinylLSTM+CNN was 20 seconds. However, the number of epoch used was $60$ and the batch size is $256$. This meant that for $60$ epochs, the average running time difference was $540$ seconds, and for $256$ batch size, the average running time difference was $38.4$ hours, using the same computing resource, i.e., a Intel core i7 HP computer. Although, this running time difference will be reduced by using a graphics processing unit (GPU). However, when the size of the datasets increase, the running time difference will become obvious.

There are other traditional machine learning methods for predicting succinylation sites that utilizes other features apart from the amino-acid sequence. Such features include the physiochemical properties such as the Pseudo Amino Acid Composition (PAAC), K-spaced Amino Acid Pairs (AAP) etc. These machine learning methods are iSuc-PseAAC \cite{xu2015isuc}, iSuc-PseOpt \cite{jia2016isuc}, PSuc-Lys \cite{jia2016psuc}, SuccinSite \cite{hasan2016succinsite},SuccinSite2.0 \cite{hasan2017systematic} GPSuc\cite{hasan2018gpsuc}
and PSuccE \cite{ning2018detecting}. Table III shows that the performance of these machine learning methods in comparison with DeepSuccinylLSTM+CNN. The values for sensitivity, specificity and MCC for the machine learning methods were obtained directly from the published articles. It can be seen from Table III that DeepSuccinylCNN+LSTM still performs better than these machine learning methods. Hence, DeepSuccinylCNN+LSTM provides a good trade-off between speed of computation and performance.

\begin{table*}
\caption{Traditional machine learning Versus DeepSuccinylCNN+LSTM}
\centering
    \begin{tabular}{c |c| c |c}
    \hline
Prediction Schemes  &  Sensitivity & Specificity & MCC\\
\hline
 iSuc-PseAAC &  0.12 & 0.89 & 0.01\\
 iSuc-PseOpt & 0.30 & 0.70 & 0.04\\
 Psuc-Lys & 0.22 & 0.83 & 0.04\\
 SuccinSite & 0.37 & 0.88 & 0.20\\
 SuccinSite2.0 & 0.45 &0.88 &0.26\\
 GPSuc & 0.50 & 0.88 & 0.30 \\
 PSuccE & 0.38 & 0.89 & 0.20 \\
 DeepSuccinylSiteCNN+LSTM & 0.77 & 0.63 &0.39\\

\hline
    \end{tabular}
     
    \label{table:2}
\end{table*}

\section{Brief Overview of Convolutional Neural Networks}
Convolutional neural network  (CNN) was inspired by the biological processes in the visual cortex of a rabbit. Unlike multi-layer perceptron which is fully connected, the CNN reduces the number of training parameters by sharing weights. CNN has been successfully for image classification \cite{krizhevsky2012imagenet}. It has been successful for automatically extracting features without the need for human intervention unlike traditional machine learning techniques. However, it still has the challenge of pose estimation in 3D images. 

CNN consists of an input layer, at least one hidden layer, and an output layer. The input layer interfaces with the object. The hidden layers is necessary for feature engineering. The hidden layers comprises the convolutional layer, where a sliding-dot product operation occurs, the ReLu layer as the activation function, the pooling layer for downsampling and feature extraction, and the fully connected dense layer that identifies the object. The output layer does the classification using the softmax function. 

\section{Brief Overview of Long Short Term Memory}
Long Short Term Memory (LSTM) is a type of Recurrent Neural Network (RNN) deep learning architecture useful for processing sequential data such as speech and text \cite{hochreiter1997long}.  Also, it is useful for processing time-series data. Unlike CNN which is a feedforward architecture, LSTM possesses feedback architecture.  

LSTM consists of a cell, an input gate, and a forget gate. The cell remembers values over arbitrary time intervals, while the three gates control the flow of information in and out of the cell. One advantage of LSTM over traditional RNN is that it overcomes the challenge of vanishing gradient encountered in traditional RNNs. The activation function for LSTM is commonly the sigmoid function.

\section{Conclusion and Future Work}
Succinylation is a post-translational modification process in proteins that plays an important role in many cellular functions. Finding the succinylation sites using experimental methods such as mass spectrometry is very laborious. Hence, computational methods are favored.  Therefore, 
this paper used an improved deep learning architecture called DeepSuccinylCNN+LSTM to predict positive succinylation sites. First, the performance of DeepSuccinylSiteCNN+LSTM was compared with DeepSuccinylSite, the state-of-the-art deep learning architecture for succinylation. It was shown that DeepSuccinylCNN+LSTM performed close to DeepSuccinylSite, but with a better trade-off in terms of computational speed. Second, the performance of DeepSuccinylCNN+LSTM was compared with other traditional machine learning architectures for succinylation in literature. It was shown that DeepSuccinylCNN+LSTM performed better using some well-known performance metrics. Hence, it can be concluded that DeepSuccinylCNN+LSTM is better favored when large datasize is used, judging by its computational speed and performance accuracy.

This work can be extended to finding better deep learning architecture that will have the best performance in terms of speed and other performance metrics. More so, novel embedding technique may be used for better performance.

\bibliographystyle{IEEEtran}
\bibliography{Ref}

\begin{thebibliography}{10}
\providecommand{\url}[1]{#1}
\csname url@samestyle\endcsname
\providecommand{\newblock}{\relax}
\providecommand{\bibinfo}[2]{#2}
\providecommand{\BIBentrySTDinterwordspacing}{\spaceskip=0pt\relax}
\providecommand{\BIBentryALTinterwordstretchfactor}{4}
\providecommand{\BIBentryALTinterwordspacing}{\spaceskip=\fontdimen2\font plus
\BIBentryALTinterwordstretchfactor\fontdimen3\font minus
  \fontdimen4\font\relax}
\providecommand{\BIBforeignlanguage}[2]{{%
\expandafter\ifx\csname l@#1\endcsname\relax
\typeout{** WARNING: IEEEtran.bst: No hyphenation pattern has been}%
\typeout{** loaded for the language `#1'. Using the pattern for}%
\typeout{** the default language instead.}%
\else
\language=\csname l@#1\endcsname
\fi
#2}}
\providecommand{\BIBdecl}{\relax}
\BIBdecl

\bibitem{xu2015isuc}
Y.~Xu, Y.-X. Ding, J.~Ding, Y.-H. Lei, L.-Y. Wu, and N.-Y. Deng, ``isuc-pseaac:
  predicting lysine succinylation in proteins by incorporating peptide
  position-specific propensity,'' \emph{Scientific reports}, vol.~5, p. 10184,
  2015.

\bibitem{jia2016isuc}
J.~Jia, Z.~Liu, X.~Xiao, B.~Liu, and K.-C. Chou, ``isuc-pseopt: identifying
  lysine succinylation sites in proteins by incorporating sequence-coupling
  effects into pseudo components and optimizing imbalanced training dataset,''
  \emph{Analytical biochemistry}, vol. 497, pp. 48--56, 2016.

\bibitem{jia2016psuc}
------, ``psuc-lys: predict lysine succinylation sites in proteins with pseaac
  and ensemble random forest approach,'' \emph{Journal of theoretical biology},
  vol. 394, pp. 223--230, 2016.

\bibitem{hasan2016succinsite}
M.~M. Hasan, S.~Yang, Y.~Zhou, and M.~N.~H. Mollah, ``Succinsite: a
  computational tool for the prediction of protein succinylation sites by
  exploiting the amino acid patterns and properties,'' \emph{Molecular
  bioSystems}, vol.~12, no.~3, pp. 786--795, 2016.

\bibitem{hasan2017systematic}
M.~M. Hasan, M.~S. Khatun, M.~N.~H. Mollah, C.~Yong, and D.~Guo, ``A systematic
  identification of species-specific protein succinylation sites using joint
  element features information,'' \emph{International journal of nanomedicine},
  vol.~12, p. 6303, 2017.

\bibitem{hasan2018gpsuc}
M.~M. Hasan and H.~Kurata, ``Gpsuc: Global prediction of generic and
  species-specific succinylation sites by aggregating multiple sequence
  features,'' \emph{PloS one}, vol.~13, no.~10, p. e0200283, 2018.

\bibitem{ning2018detecting}
Q.~Ning, X.~Zhao, L.~Bao, Z.~Ma, and X.~Zhao, ``Detecting succinylation sites
  from protein sequences using ensemble support vector machine,'' \emph{BMC
  bioinformatics}, vol.~19, no.~1, p. 237, 2018.

\bibitem{thapa2020deepsuccinylsite}
N.~Thapa, M.~Chaudhari, S.~McManus, K.~Roy, R.~H. Newman, H.~Saigo, and D.~B.
  Kc, ``Deepsuccinylsite: a deep learning based approach for protein
  succinylation site prediction,'' \emph{BMC bioinformatics}, vol.~21, pp.
  1--10, 2020.

\bibitem{krizhevsky2012imagenet}
A.~Krizhevsky, I.~Sutskever, and G.~E. Hinton, ``Imagenet classification with
  deep convolutional neural networks,'' in \emph{Advances in neural information
  processing systems}, 2012, pp. 1097--1105.

\bibitem{hochreiter1997long}
S.~Hochreiter and J.~Schmidhuber, ``Long short-term memory,'' \emph{Neural
  computation}, vol.~9, no.~8, pp. 1735--1780, 1997.

\end{thebibliography}

\end{document}